\documentclass[runningheads]{llncs}
\usepackage[T1]{fontenc}
\usepackage{graphicx}
\usepackage{amsmath}
\usepackage{booktabs}
\usepackage{xcolor}
\usepackage{tikz}
\usepackage{bbding}
\usepackage{multirow} 

\newcommand{\blue}[1]{\textcolor{blue!80!black}{#1}}

\begin{document}
\title{Integrating of Attack Patterns and Threat Modeling to Support Security Analysis in the Automotive Domain}
\author{Eduardo S. O. Marques\Envelope\inst{1}\orcidID{0000-0002-5503-5302} \and
Saulo Moraes Villela\inst{1,2}\orcidID{0000-0001-5958-4766} \and
Heder Soares Bernardino\inst{1,2}\orcidID{0000-0003-2012-7802} \and
Alex Borges Vieira\inst{1,2}\orcidID{0000-0003-0821-126X} \and
Bernardo Martins Rocha\inst{1,2}\orcidID{0000-0002-0508-8959} \and
André Luiz de Oliveira\inst{1,2}\orcidID{0000-0003-0564-0034}}
\authorrunning{E. S. O. Marques et al.}
\institute{Graduate Program in Computational Modeling, Federal University of Juiz de Fora, Juiz de Fora, Brazil \\ \email{eduardo.santos@estudante.ufjf.br} \and
Department of Computer Science, Federal University of Juiz de Fora, Juiz de Fora, Brazil \\
\email{\{saulo.moraes,heder.bernardino,alex.borges,	
bernardomartinsrocha,andre.oliveira\}@ufjf.br}}
\maketitle              
\pagestyle{empty} 
\begin{abstract}
The increasing connectivity of cyber-physical systems in the automotive domain demands systematic approaches for identifying and mitigating cybersecurity threats. While ISO 21434 defines a structured workflow for Threat Analysis and Risk Assessment (TARA), it does not prescribe a specific method for identifying attack paths or evaluating risks. This paper proposes a conceptual model grounded in ISO 21434 that integrates the STRIDE threat classification with the CAPEC attack pattern database, focusing on the identification of attack scenarios and risk assessment. Its applicability is demonstrated through a case study involving Bluetooth-based threats to an automotive headlamp system.

\end{abstract}
%
\section{Introduction}

Autonomous vehicles represent a significant evolution in the automotive sector, characterized as cyber-physical systems capable of making decisions and performing actions with minimal human intervention. To reach this level of autonomy, these vehicles need to continuously interact with other vehicles and with road infrastructure, such as traffic lights and signs \cite{aslansefat2021toward}. As critical systems, any functional failure or cyberattack can compromise not only system performance but also the safety and lives of occupants and the surrounding environment \cite{biro2017software}.
 
The growing interconnectivity and advancement of autonomous vehicles introduce significant cybersecurity challenges, increasing the attack surface of these systems.
In order to mitigate these risks, the automotive sector has adopted standards such as ISO/SAE 21434 \cite{standard2021iso}, which defines an approach based on threat analysis and risk assessment, known as TARA (\textit{Threat Analysis and Risk Assessment}). Although
the standard does not prescribe specific methods for carrying out the analysis, this approach proposes the use of risk models at different levels of system abstraction. At the system level, models such as STRIDE (\textit{Spoofing, Tampering, Repudiation, Information Disclosure, Denial of Service, Elevation of Privilege}) can be employed to identify threat scenarios and damage associated with critical assets \cite{shostack2014threat}. At the component level, methods such as Attack Tree Analysis allow the description of possible paths for carrying out attacks \cite{schneier1999attack}, while attack vector models help classify risks according to the exploitable surface.
 
Attack catalogs, such as CAPEC (\textit{Common Attack Pattern Enumeration and Classification}) \cite{mitre2023capec}, are fundamental tools to support the identification and classification of risks. When combined with STRIDE, they can provide valuable support for attack path analysis. However, despite the relevance of these tools, there is still a gap in the literature regarding the structured integration between threat analysis models and databases in the automotive context. Many studies focus on isolated aspects of risk analysis, without exploring the potential synergy between models such as STRIDE and catalogs such as CAPEC in supporting the TARA process.
 
Given this scenario, this work proposes a methodology that integrates the STRIDE model with the CAPEC catalog to support the identification of threat scenarios, damage, and attack paths in compliance with the ISO 21434 standard. The proposal aims to fill the integration gap between risk models and structured databases, providing a more systematic and well-founded process for threat analysis in automotive systems. As a scientific contribution, this work proposes (i) the development of a structured database to store and relate information extracted from the STRIDE-CAPEC bases, (ii) a methodological approach that systematizes the joint use of STRIDE and CAPEC in the TARA process; and (iii) an evaluation through a case study based on examples from ISO 21434 itself, demonstrating the feasibility and benefits of the proposed methodology. The results show that the integration between models and catalogs contributes to a more comprehensive and accurate identification of attack paths and risks associated with autonomous automotive systems.
 
The remainder of this paper is organized as follows: Section~\ref{sec:theory} presents the theoretical background and the concepts used; Section~\ref{sec:contribution} describes the proposed methodology; Section~\ref{sec:example} defines the application example extracted from the standard; Section~\ref{sec:results} presents the experimental evaluation and the results obtained; and, finally, Section~\ref{sec:conclusion} discusses the conclusions and future perspectives of this work.

\section{Background} \label{sec:theory}

In order to conceptually ground this work, this section presents the theoretical framework necessary to understand the foundations used in cybersecurity risk modeling in the automotive context. First, the main concepts of the ISO/SAE 21434 standard, which establishes guidelines for cybersecurity engineering in road vehicles, are discussed. Next, the STRIDE model, used for threat categorization, is addressed, and finally, the CAPEC database, which offers a detailed taxonomy of attack techniques. The integration of these elements constitutes the methodological basis of the threat analysis and of the proposal developed in this study.

\subsection{ISO 21434} 
 
\begin{figure}[ht]
\centering
\includegraphics[width=1.0\textwidth, page=1, trim={5.2cm 22.0cm 5.2cm 1.6cm},clip]{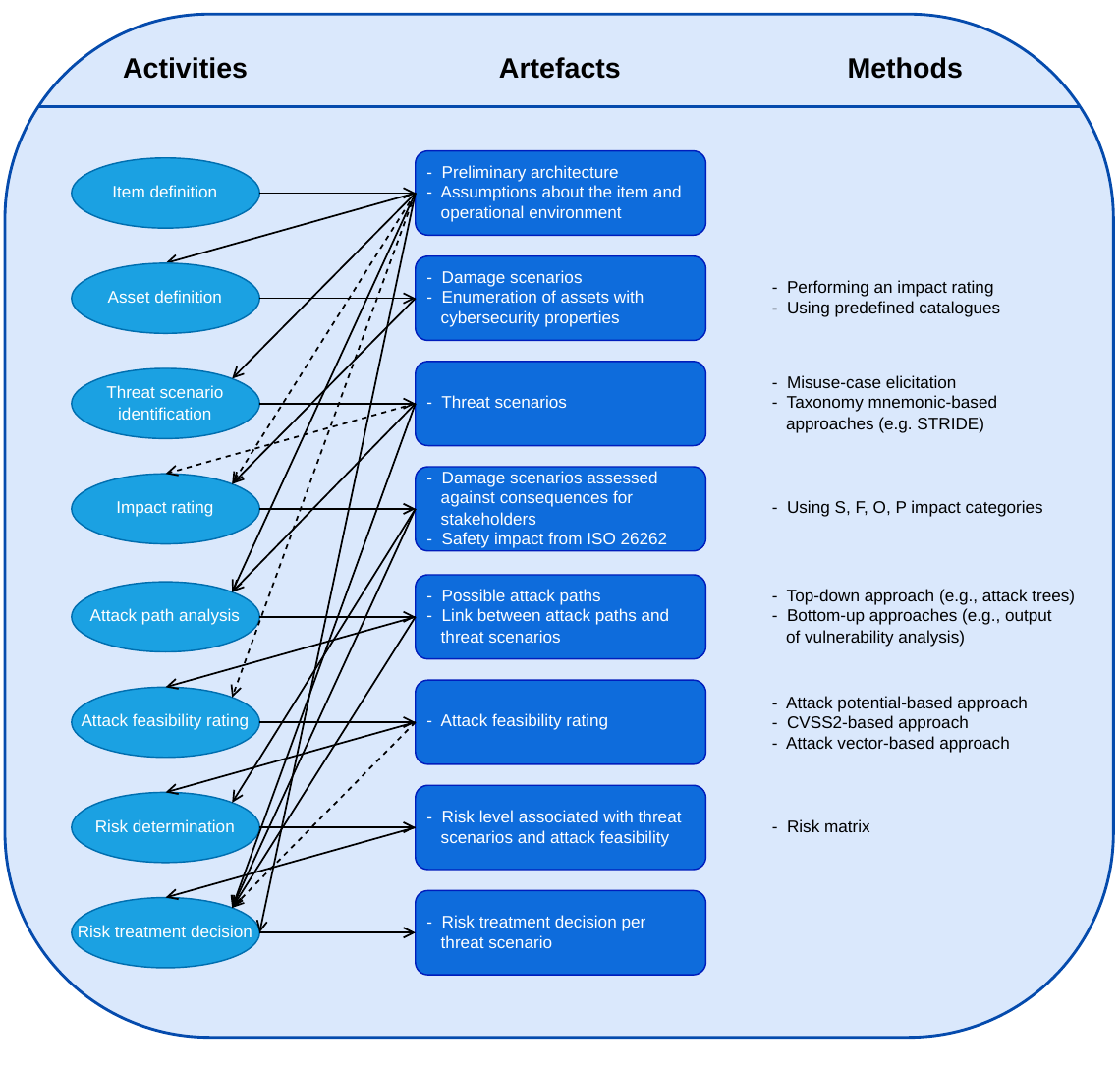}
\caption{Relationship between terms defined in ISO 21434 (adapted from  \cite{standard2021iso}).}
\label{fig:ISO_steps}
\end{figure}

ISO/SAE 21434 Road Vehicles - Cybersecurity Engineering \cite{standard2021iso} is an automotive industry standard that defines cybersecurity engineering requirements for road vehicles, focusing on the application of a risk-based approach. Its main objective is to ensure digital security throughout the entire vehicle lifecycle, from concept to decommissioning, promoting the systematic identification, assessment, and mitigation of cyber threats and risks. However, the standard does not prescribe specific technical methods or ready-made solutions, nor does it define exclusive requirements for autonomous vehicles or road infrastructure. Instead, it proposes a structured analysis workflow that can be adapted according to the system context and the threats identified \cite{macher2020iso}. Figure~\ref{fig:ISO_steps} illustrates this structured workflow proposed by ISO 21434 for conducting TARA, highlighting the relationship between activities, generated artifacts, and recommended methods at each step.
 
The process begins with the definition of the item (system), which includes the definition of the preliminary architecture, as well as relevant assumptions about the item and its operational environment. In ISO, an \textbf{item} represents a component or a set of components that implement a function at the vehicle level. After defining the boundaries of the item and its operational environment, the assets, their associated cybersecurity properties, and damage scenarios are identified. An \textbf{asset} is an object that has value or contributes to a value, and can be a system, a component, data (information), or data communication. A \textbf{cybersecurity property} is an attribute of an asset whose protection is considered essential, such as confidentiality, integrity, and/or availability. A \textbf{damage scenario} is an adverse consequence involving a vehicle or its function, affecting a road user. A \textbf{threat scenario} is a potential cause of a breach of the cybersecurity properties of one or more assets that realizes a damage scenario.

The \textbf{asset identification} activity involves the \textbf{damage scenario identification} step, whose objective is to recognize possible violations of the cybersecurity properties associated with each asset. The damage scenarios identified are then assessed regarding their possible adverse consequences for road users, categorizing the impact along the following dimensions: safety, financial, operational, and privacy (S, F, O, P). \textbf{Impact} refers to the degree of severity of the damage or physical harm resulting from the realization of a damage scenario. The \textbf{impact rating} can be defined as severe, major, moderate, or negligible, according to the criteria established for the assessment, which are derived from the ISO 26262 functional safety standard \cite{iso2018road}, applicable to road vehicles.
 
\textbf{Threat scenario identification} describes a set of potential actions that can lead to one or more damage scenarios, which specify the adverse consequences (outcomes) of an attack, for example, using the STRIDE threat analysis method. A threat scenario can lead to the occurrence of several damage scenarios. ISO 21434 also prescribes conducting attack path analysis to identify the possible attacks that can realize each threat scenario. An \textbf{attack path} is a set of deliberate actions to realize a threat scenario. Attack path analysis can be conducted using \textit{top-down} approaches, starting from the threat scenario to identify possible paths, or \textit{bottom-up} approaches, which start from vulnerabilities detected in system assets to infer possible threats and attack routes. Subsequently, each identified path must receive an attack feasibility rating (high, medium, low, or very low). The \textbf{attack feasibility rating} can be determined based on criteria such as required time, prior knowledge, window of opportunity, and technical resources, as defined by ISO/IEC 18045 \cite{iso2008iso}. The standard also defines an approach for setting cybersecurity goals for each identified threat scenario in the form of a CAL. Cybersecurity goals define the applicable risk-sharing, prevention, reduction, or acceptance measures, according to the threat level associated with the asset. ISO 21434 also recommends producing a cybersecurity case to provide the argument for the cybersecurity of the item or component, supported by work products.

\subsection{STRIDE}
 
\begin{figure}[ht]
\centering
\includegraphics[width=1.0\textwidth, page=1, trim={5.5cm 34.5cm 5.5cm .8cm},clip]{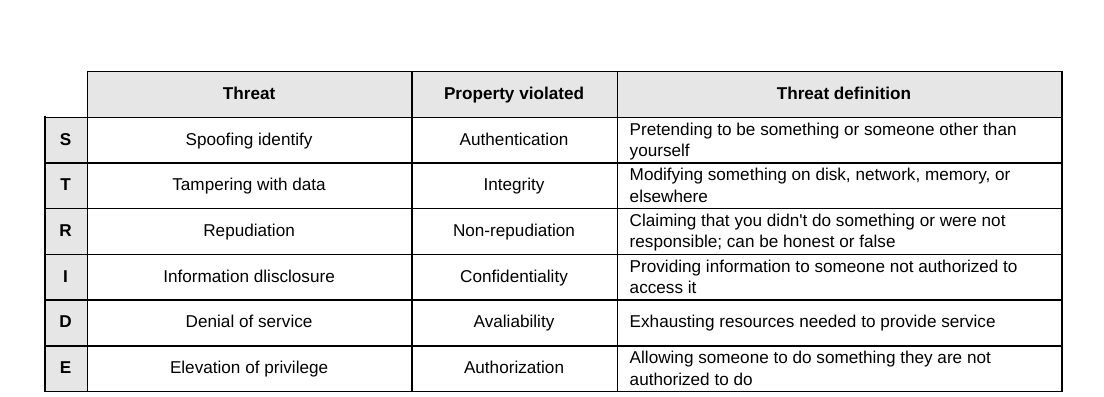}
\caption{STRIDE threat categories (adapted from \cite{shevchenko2018threat}).}
\label{fig:STRIDE}
\end{figure}

STRIDE is a methodology proposed by Microsoft that is widely used to identify and categorize cyber threats in computer systems \cite{swiderski2004threat}. The term STRIDE is derived from the initial letters of different threats, classified based on the objectives and purposes of the attacks. The threats are: spoofing, tampering, repudiation, information disclosure, denial of service (DoS), and elevation of privilege. Figure~\ref{fig:STRIDE} illustrates these categories and their relationship with the cybersecurity properties typically violated in computer systems. This model allows security engineers to analyze potential vulnerabilities in complex systems, providing a systematic framework for threat modeling \cite{shostack2014threat}.
 
The STRIDE methodology allows for a systematic view of cybersecurity risks by identifying threat scenarios based on the analysis of interactions between system elements and how they can be exploited \cite{ward2021automotive}. Each STRIDE category can encompass multiple threats, just as a single threat can be associated with more than one category \cite{tatam2021review}, which contributes to a more effective organization of the security strategy, with planned responses for different types of attacks. However, the model does not directly address the level of detail or specific attack methods \cite{bodeau2018cyber}, being limited to the general categorization of threat types. Thus, to ensure a more complete analysis, it is essential to consider data sources that describe adversarial behaviors, exploitation techniques, and known vulnerabilities.

\subsection{Attack Catalog: CAPEC}
 
\begin{figure}[ht]
\centering
\includegraphics[width=1.0\textwidth]{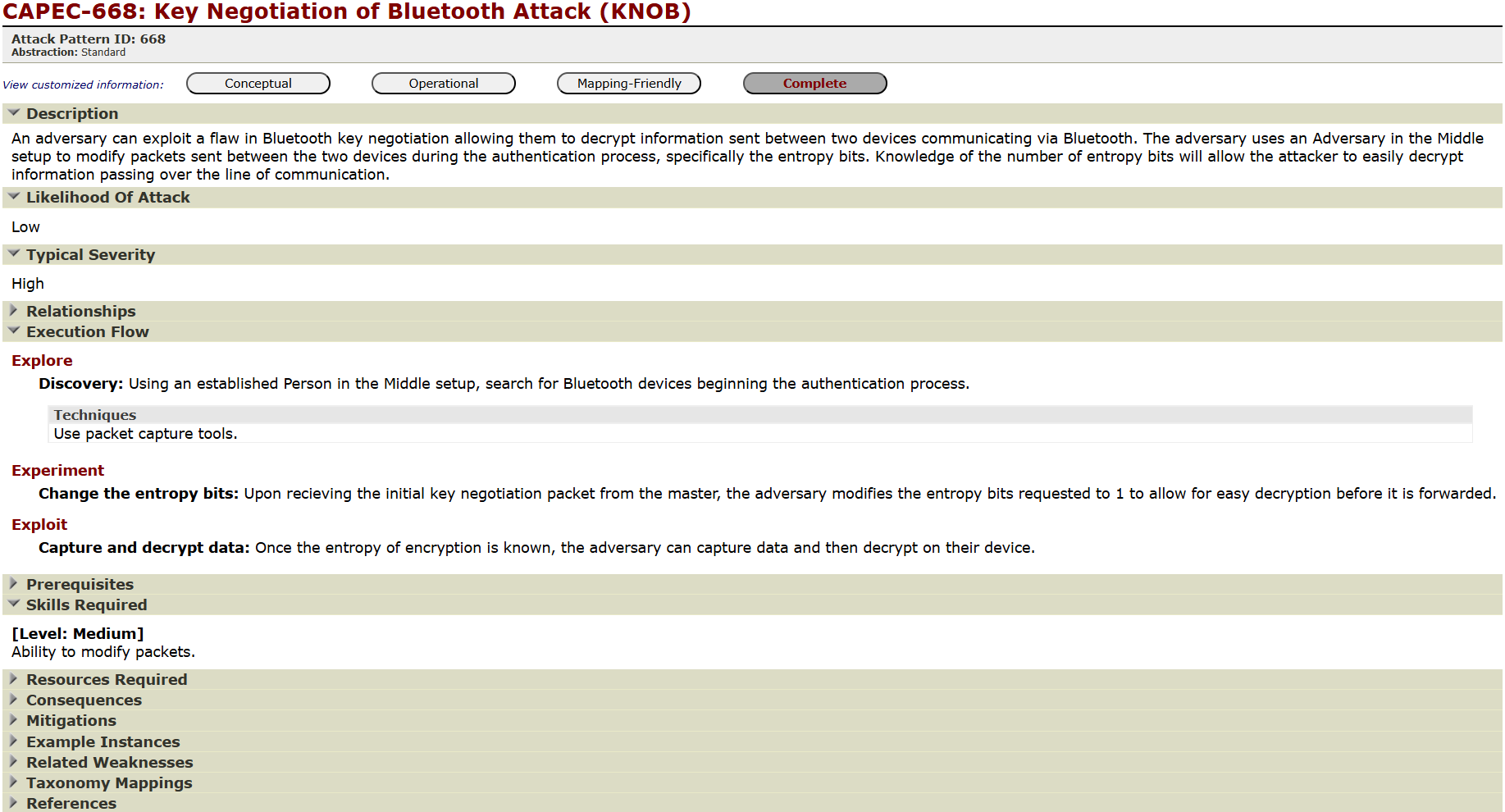}
\caption{Example of a CAPEC entry \cite{mitre2023capec}.}
\label{fig:CAPEC_example}
\end{figure}

The \textit{Common Attack Pattern Enumeration and Classification} (CAPEC) is an online catalog of attack patterns, containing more than 500 entries. Maintained by the MITRE Corporation, CAPEC provides detailed information on techniques, methods, and strategies used by malicious agents to exploit vulnerabilities \cite{mitre2023capec}. CAPEC can be used to help create effective countermeasures and develop secure architectures. An example of a CAPEC entry is shown in Figure~\ref{fig:CAPEC_example}.
 
Each attack pattern in CAPEC is described through several attributes that can be used to help understand and mitigate threats. Among these attributes, the following stand out: likelihood of attack, severity, consequences, required skills, weaknesses, and related attack patterns. An attack catalog focuses on the behavior of attackers and the methods employed to compromise systems. This allows security professionals to better understand the lifecycle of an attack and implement strategies for prevention and response.

\section{Contributions} \label{sec:contribution} 
 
This section presents the main contributions of this work, which include the construction of a database dedicated to the structured storage of information extracted from the CAPEC attack pattern catalog, as well as the development of a methodological approach to systematically integrate the STRIDE and CAPEC models into the TARA process in compliance with ISO/SAE 21434.

\subsection{Database: CAPEC and STRIDE}
 
To support risk analysis in compliance with ISO/SAE 21434, this work proposes the construction of a document-oriented database, using NoSQL through MongoDB\footnote{\blue{\url{https://www.mongodb.com/docs/}}}. The main objective was to systematically integrate the data from the CAPEC and STRIDE catalogs, enabling a more contextualized and effective identification of cyber threats. CAPEC information was obtained directly from the official MITRE database (https://capec.mitre.org/data/downloads.html), ensuring the use of up-to-date attack pattern descriptions. For STRIDE, a pre-established mapping was used based on the CAPEC-STRIDE Mapping project \cite{OSTERING}, in which each CAPEC pattern is associated with one or more categories of the STRIDE model \cite{shostack2014threat}.
 
The database structuring was carried out with the help of the DbSchema tool\footnote{\blue{\url{https://dbschema.com}}}, which enables graphical visualization of the logical model. This approach facilitated the mapping of existing connections in the attack catalog, such as severity, likelihood, and complexity. The tool was also fundamental for ensuring the integrity of the relationships and the documentation of the model, contributing to the scalability and maintainability of the solution. Furthermore, it allowed visual validation of the system's logical model, ensuring adherence to the conceptual workflow proposed by ISO/SAE 21434.
 
\begin{figure}[ht]
\centering
\includegraphics[width=0.8\textwidth, page=1, trim={3.2cm 15.3cm 4.1cm 1.4cm},clip]{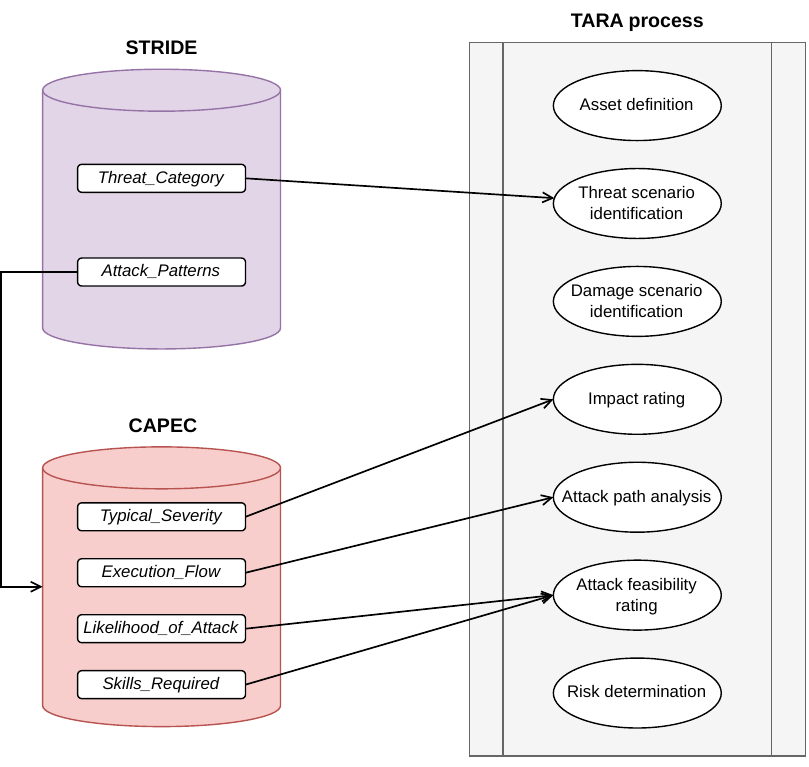}
\caption{Relationship between the database and ISO TARA 21434.}
\label{fig:Database_Relation}
\end{figure}

The main contribution of this approach lies in the systematization of the relationship between the STRIDE and CAPEC models, allowing attack patterns to be categorized according to the threat's objectives. Unlike the work proposed by Crawley \cite{OSTERING}, which presents a manual and descriptive, static mapping between CAPEC and STRIDE, this work proposes a reusable and scalable relational structure, capable of integrating new threats and additional attributes relevant to risk assessment. Through this integration, it is possible to enrich the threat modeling process, supporting activities such as attack scenario identification, exploitation path analysis, and attack feasibility assessment. Figure~\ref{fig:Database_Relation} presents the relationship between the database entities and the steps of the TARA process defined by ISO/SAE 21434 (see the Appendix at the end of the document).
 
\subsection{Integration with ISO 21434}
 
\begin{figure}[ht]
\centering
\includegraphics[width=1.0\textwidth, page=1, trim={0.8cm 11.0cm 2.8cm 1.0cm},clip]{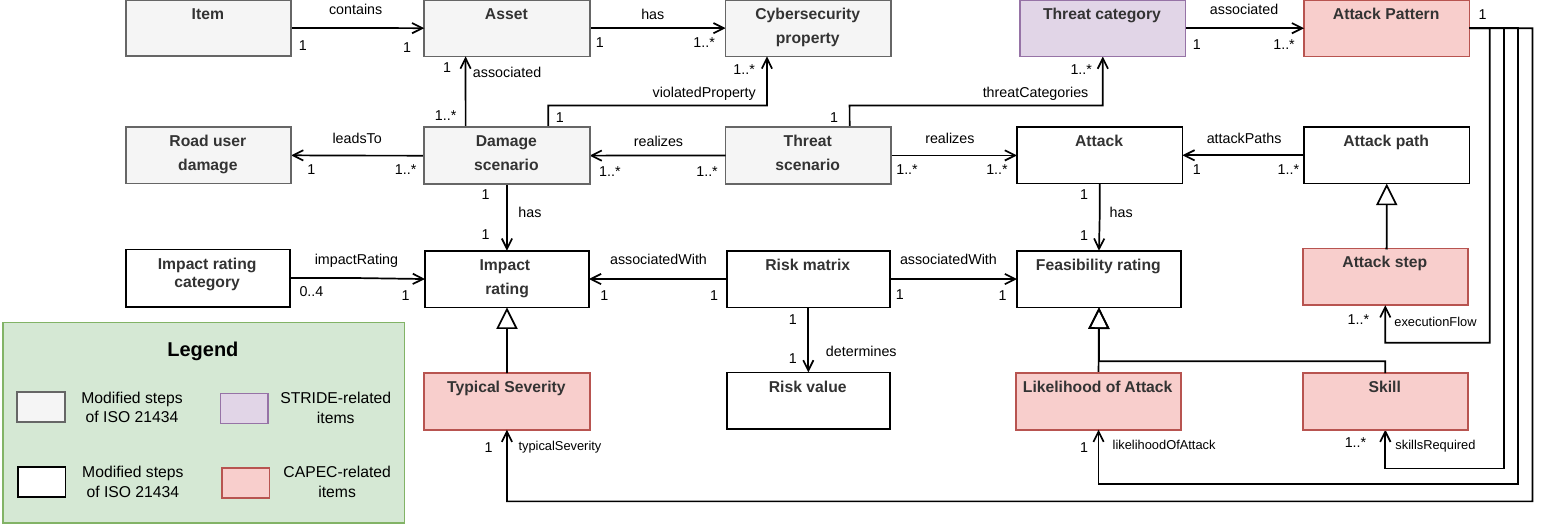}
\caption{Ontology integrating ISO TARA 21434 with STRIDE-CAPEC.}
\label{fig:Model_Database}
\end{figure}

In order to organize and structure threat analysis in autonomous automotive systems, an ontology was developed that establishes relationships between the main security elements defined by ISO/SAE 21434 and the STRIDE and CAPEC catalogs. This model, illustrated in Figure~\ref{fig:Model_Database}, provides a systematic approach for threat identification, impact assessment, and risk value determination, promoting greater consistency and traceability in the TARA process.
 
The model preserves the initial steps proposed by ISO 21434, starting with the definition of items and assets, followed by the identification of threat and damage scenarios. From there, the notion of \textbf{attacks} is introduced, which materialize from threat scenarios and unfold into \textbf{attack paths}. Each path is composed of a sequence of \textbf{attack steps}, modeled from the execution flow described in the CAPEC patterns, allowing the exploitation logic of vulnerabilities to be represented with precision.
 
The feasibility of each attack is analyzed through the feasibility rating, which considers two central attributes: the \textbf{likelihood of attack occurrence} and \textbf{expertise} (\textit{skill}). These entities, extracted from CAPEC, allow estimating the effort required to execute an attack, supporting risk prioritization and the definition of countermeasures.
 
Furthermore, the model addresses the assessment of impacts caused by successful attacks. Damage scenarios are associated with the possible effects on vehicle occupants and third parties, categorized according to the impact rating. This rating is determined based on the \textbf{severity} of the related attack patterns, as indicated in CAPEC. In this way, the model allows the potential harm to be estimated and integrated directly into the risk matrix, from which the risk value is obtained.
 
This integration between normative elements and technical databases allows for a more robust and contextualized analysis, connecting the conceptual modeling of ISO 21434 with concrete sources of attack patterns and threats. The result is a framework that favors the practical application of TARA in automotive systems, with greater depth and precision in cybersecurity risk analysis.

\section{Case Study} \label{sec:example} 
 
To evaluate the feasibility of the proposed model, two examples extracted from Annex H of ISO 21434 will be used, representing Bluetooth vulnerabilities in a headlamp control system for spoofing and denial-of-service (DoS) attack scenarios. The choice of this scenario is justified by the presence of the Bluetooth protocol as a means of communication between components, which makes it particularly relevant for threat analysis in cyber-physical systems. In this context, wireless connectivity introduces additional attack vectors, increasing the exposure of security vulnerabilities.

\subsection{Preliminary Architecture: Headlamp System}
 
Below, a conceptual and simplified view of the headlamp system is presented. This representation highlights the boundaries of the item, as well as its components and their interaction with the operating system, serving as a basis for the threat analysis in the selected scenarios.
 
\begin{figure}[ht]
\centering
\includegraphics[width=0.9\textwidth, page=1, trim={4cm 26cm 4cm 0.4cm},clip]{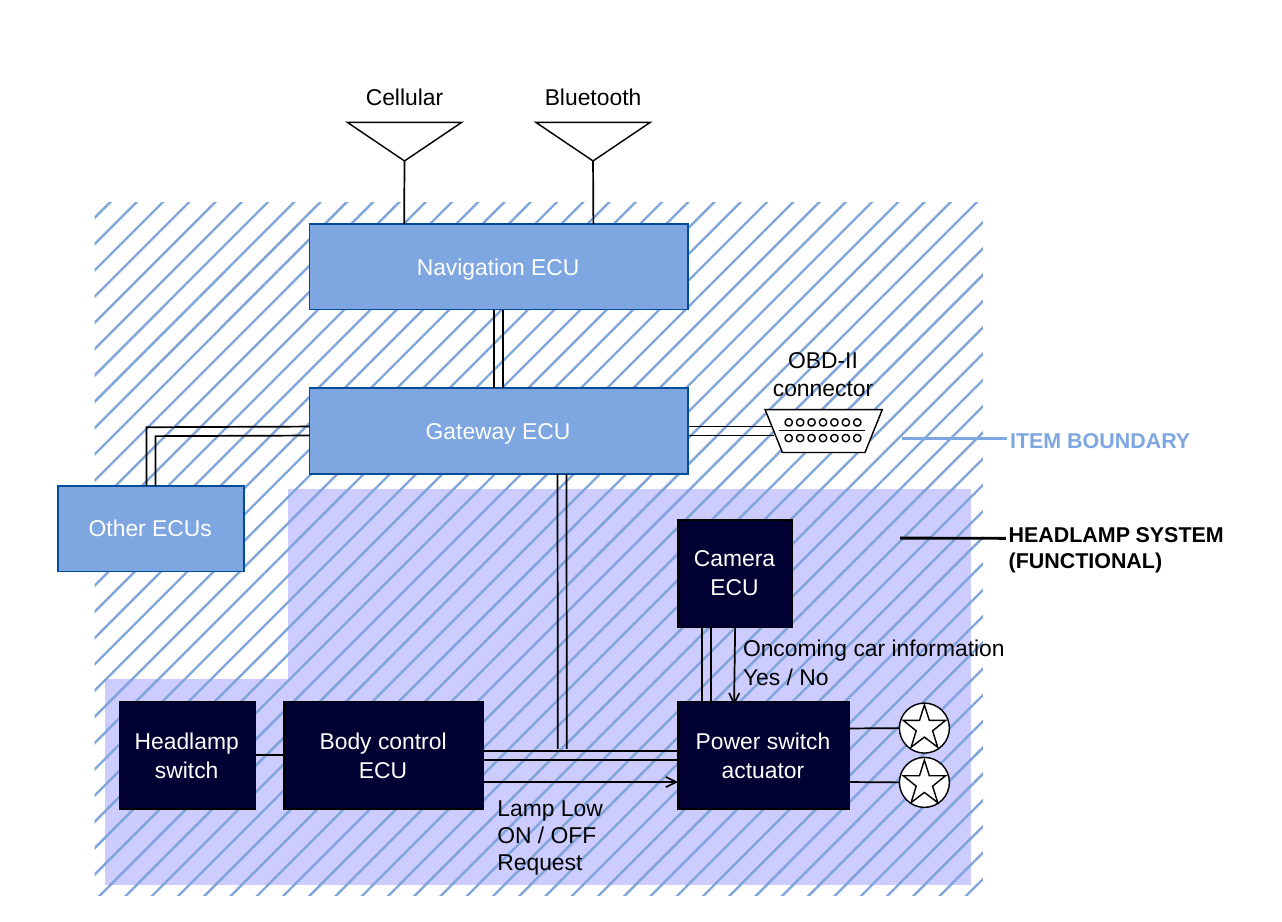}
\caption{Headlamp system architecture (adapted from  \cite{dantas2020security}).}
\label{fig:Headlamp}
\end{figure}

A headlamp system is responsible for turning the headlamp on/off according to the driver's demand. This system also has two light modes when turned on: high beam and low beam. If the light is in high-beam mode, the system automatically switches to low-beam mode when an approaching vehicle is detected by the camera. It automatically returns to high-beam mode if no more vehicles are detected. The headlamp system is a critical system, as accidents can occur if the lamp is unexpectedly turned off at night. The preliminary architecture adopted in this work is illustrated in Figure~\ref{fig:Headlamp}.
 
The body control ECU (\textit{Electronic Control Unit}) transmits messages to the power switch actuator via the CAN (\textit{Controller Area Network}) bus, sending commands to activate and deactivate the headlamps. To automatically switch between high- and low-beam modes, a camera ECU is integrated with the power switch actuator, enabling the detection of approaching vehicles. The CAN bus is connected to a gateway ECU, responsible for regulating the access of other external ECUs to the headlamp system, such as the navigation ECU. The navigation ECU has two communication interfaces: cellular and Bluetooth. In addition, the gateway ECU is connected to an OBD-II connector via a separate CAN bus. OBD-II is an on-board diagnostic module that collects and monitors vehicle data, including speed. The Bluetooth, cellular, and OBD-II interfaces lie outside the item's boundary and can serve as possible attack vectors on the headlamp system, as discussed below.

\subsection{Application Example of the TARA Method}
 
After defining the item and its operational environment, the TARA method is applied to evaluate the risk matrix. For the presentation of this paper, the method will be carried out as follows: damage scenario identification; threat scenario identification; attack path analysis; attack feasibility rating; impact rating; and risk value determination.
 
Table~\ref{tab:assets} presents the identification of the relevant assets and their respective damage scenarios, relating each asset to the affected cybersecurity properties — confidentiality (C), integrity (I), and availability (A). Table~\ref{tab:threats} expands this analysis by describing the corresponding threat scenarios, connecting potential damage to possible causes, namely signal spoofing and denial of service. Finally, Table~\ref{tab:attack_path} details the attack paths associated with each threat scenario, revealing the sequence of compromises that can occur across a chain of ECUs, exploiting the Bluetooth interface as the initial vector.
 
\begin{table}[ht]
    \centering
    \caption{List of assets and damage scenarios (adapted from \cite{standard2021iso}).}
    \begin{tabular}{|p{4cm}|c|c|c|p{8cm}|}
        \hline
        \textbf{Asset} & \multicolumn{3}{c|}{\textbf{Cyb. Prop.}} & \textbf{Damage scenario} \\
        \cline{2-4}
        & \textbf{C} & \textbf{I} & \textbf{A} & \\
        \hline
        Data communication (lamp request) 
        & -- & X & -- & Frontal collision with a narrow, fixed object (e.g., a tree) caused by unintended headlamp shutdown while driving at night at medium speed \\
        \hline
        Data communication (approaching car information)
        & -- & -- & X & Malfunction of the automatic high beam caused by the headlamp permanently remaining in low-beam mode while driving at night \\
        \hline
    \end{tabular}
    \label{tab:assets}
\end{table}
 
\begin{table}[ht]
    \centering
    \caption{Threat scenarios associated with the damage (adapted from \cite{standard2021iso}).}
    \begin{tabular}{|p{4cm}|p{10cm}|}
        \hline
         \textbf{Damage scenario} & \textbf{Threat scenario} \\
        \hline
        Frontal collision with [...] & Spoofing of a signal leads to loss of integrity of the data communication of the lamp request signal to the power switch actuator, which may cause the headlamp to shut down unintentionally \\
        \hline
         Malfunction [...] & Denial of service of information about approaching cars \\
        \hline
    \end{tabular}
    \label{tab:threats}
\end{table}
 
\begin{table}[ht]
    \centering
    \caption{Attack paths for threat scenarios (adapted from \cite{standard2021iso}).}
    \begin{tabular}{|p{1.7cm}|p{12cm}|}
        \hline
        \textbf{Threat scenario} & \textbf{Attack path} \\
        \hline
        \multirow{4}{*}{Spoofing} & i. The attacker compromises the navigation ECU via the Bluetooth interface \\
        & ii. The compromised navigation ECU transmits malicious control signals \\
        & iii. The gateway ECU forwards malicious signals to the actuator \\
        & iv. Malicious signals spoof the lamp request (OFF) \\
        \hline
        \multirow{5}{*}{DoS} & i. The attacker connects a Bluetooth-enabled OBD dongle to the OBD connector \\
        & ii. The attacker compromises the driver's smartphone via the Bluetooth interface \\
        & iii. Sends a message via the smartphone and Bluetooth dongle to the gateway ECU \\
        & iv. The gateway ECU forwards malicious signals to the actuator \\
        & v. The attacker floods the communication with a large number of messages \\
        \hline
    \end{tabular}
    \label{tab:attack_path}
\end{table}
 
Table~\ref{tab:feasibility} presents the attack feasibility rating using two approaches: the attack vector-based approach and the attack potential-based approach. The first considers how remote (logically or physically) an attacker can be to exploit an attack path — the more remote, the higher the feasibility assigned. The attack potential, as defined by ISO/IEC 18045 \cite{iso2008iso}, measures the effort required to carry out an attack, based on five factors: time, expertise, knowledge of the target, window of opportunity, and equipment. In this paper, only the final result of the mapping of these parameters is presented.
 
\begin{table}[ht]
    \centering
    \caption{Attack feasibility rating (adapted from \cite{standard2021iso}).}
    \begin{tabular}{|p{8cm}|c|}
        \hline
         \textbf{Attack path} & \textbf{Attack feas. rating} \\
        \hline
        Malicious signals spoof the headlamp request & Medium \\
        \hline
        The attacker floods the communication bus & Low \\
        \hline
    \end{tabular}
    \label{tab:feasibility}
\end{table}
 
Table~\ref{tab:impact_rating} presents the impact rating across the following categories: safety, financial, operational, and privacy. The criteria for the safety category follow ISO 26262-3:2018 \cite{iso2018road}; and those related to privacy, such as personal data (PII), are defined according to ISO/IEC 29100 \cite{international2011iso}. In this paper, only the final result of the categories and their respective ratings is presented. Finally, the risk matrix is defined from the results of the attack feasibility rating and the impact rating, shown in Table~\ref{tab:risk_matrix}. The result for the headlamp example is presented in Table~\ref{tab:risk_values}.
 
\begin{table}[ht]
    \centering
    \caption{Impact rating (adapted from \cite{standard2021iso}).}
    \begin{tabular}{|p{4cm}|c|c|}
        \hline
         \textbf{Damage scenario} & \textbf{Impact cat.} & \textbf{Impact rating} \\
        \hline
        Frontal collision with [...] & S & Severe (S3) \\
        \hline
        Malfunction [...] & O & Moderate \\
        \hline
    \end{tabular}
    \label{tab:impact_rating}
\end{table}
 
\begin{table}[ht]
    \centering
    \caption{Risk matrix (adapted from \cite{standard2021iso}).}
    \begin{tabular}{|l|c|c|c|c|c|}
        \hline
        \multicolumn{2}{|c|}{} & \multicolumn{4}{c|}{\textbf{Attack feas. rating}} \\
        \cline{3-6}
        \multicolumn{2}{|c|}{} & Very low & Low & Medium & High \\
        \hline
        \multirow{4}{*}{\textbf{Impact rating}} 
        & Severe & 2 & 3 & 4 & 5 \\
        \cline{2-6}
        & Major & 1 & 2 & 3 & 4 \\
        \cline{2-6}
        & Moderate & 1 & 2 & 2 & 3 \\
        \cline{2-6}
        & Negligible & 1 & 1 & 1 & 1 \\
        \hline
    \end{tabular}
    \label{tab:risk_matrix}
\end{table}
 
\begin{table}[ht]
    \centering
    \caption{Determination of risk values (adapted from \cite{standard2021iso}).}
    \begin{tabular}{|c|c|c|c|}
        \hline
         \textbf{Threat scen.} & \textbf{Aggregated attack feas. rating} & \textbf{Impact rating} & \textbf{Risk value} \\
        \hline
        Spoofing & Medium & Severe & S : 4 \\
        \hline
        Denial of Service & Low & Moderate & O : 2 \\
        \hline
    \end{tabular}
    \label{tab:risk_values}
\end{table}

\section{Evaluation of the Proposed Model} \label{sec:results} 
 
To validate the effectiveness of the proposed model, an evaluation was conducted using the headlamp control system example presented earlier. The process followed a structured approach, starting with threat analysis using the STRIDE methodology, enabling the classification of threats according to their respective security categories. Based on this identification, an effort was made to establish the relationship between the attack vectors and the patterns documented in the CAPEC database, allowing for a structured mapping of the threats to the system.  
 
\begin{figure}[ht]
\centering
\includegraphics[width=1.0\textwidth, page=1, trim={0.8cm 27.cm 0.8cm .4cm},clip]{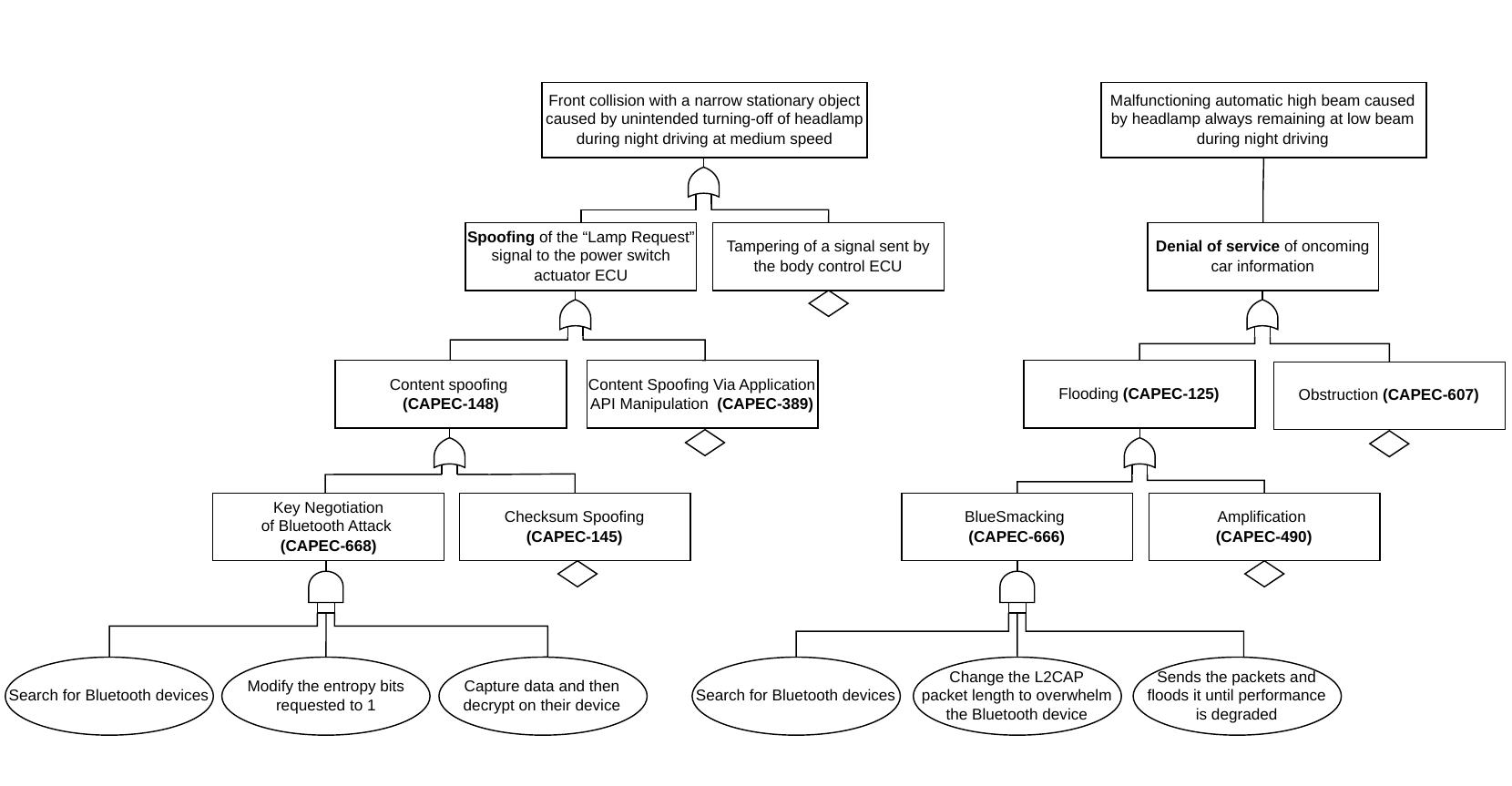}
\caption{Attack tree defined through the CAPEC catalog.}
\label{fig:tree_CAPEC}
\end{figure}

Given the context of the evaluated system, threats that compromise the integrity and availability of wireless communication via Bluetooth were identified. These threats were classified into two STRIDE categories: \textit{Spoofing} and \textit{Denial of Service}. Based on this categorization, CAPEC attack patterns that exploit Bluetooth as an attack vector were mapped. Figure~\ref{fig:tree_CAPEC} presents the attack tree built based on the identified exploitation paths.
 
Among the attacks that perform signal spoofing, CAPEC-668 (Key Negotiation of Bluetooth Attack - KNOB) stands out, which exploits key negotiation in the Bluetooth protocol, allowing an attacker to weaken the encryption and impersonate a legitimate device. For denial of service, CAPEC-666 (BlueSmacking) was identified, in which overloading the Bluetooth stack causes system unavailability. The selection of these CAPECs was based on the relationship between the threats classified in STRIDE and the documented attack patterns, ensuring the coherence of the adopted approach. The final validation of these attacks was carried out through the risk matrix, in which the feasibility and impact attributes of the attacks were analyzed according to ISO 21434 guidelines. 


\subsection{Risk Matrix Definition}
 
The attack feasibility rating for the proposed model is determined based on two attributes defined in the CAPEC taxonomy: \textit{Likelihood\_of\_Attack} and \textit{Skills\_Required}. Both attributes are categorized into three levels — low, medium, and high — to which numerical values are assigned (1 for low, 3 for medium, and 5 for high). The combination of these two parameters is used to estimate the attack feasibility rating, calculated by summing the values assigned to each attribute. This approach is based on an adaptation of the methodology proposed by the ISO/IEC 18045 standard \cite{iso2008iso}, as shown in Table~\ref{tab:likelihood_skills}.
 
\begin{table}[ht]
    \centering
    \caption{Mapping of the attack feasibility rating.}
    \begin{tabular}{|c|c|}
        \hline
         \textbf{Attack feas. rating} & \textbf{Values} \\
        \hline
        Unlikely & 0 - 2 \\
        \hline
        Possible & 3 - 6 \\
        \hline
        Likely & $>$ 6 \\
        \hline
    \end{tabular}
    \label{tab:likelihood_skills}
\end{table}
 
This categorization allows assessing how feasible it is to carry out an attack, based on the severity and likelihood of exploitation. After determining the attack feasibility, the risk value is obtained by crossing this value with the impact rating, through the \textit{Typical\_Severity} attribute, as defined in the risk matrix presented in Table~\ref{tab:risk_model}.
 
\begin{table}[ht]
    \centering
    \caption{\textbf{Definition of the risk matrix for the proposed model.}}
    \begin{tabular}{|l|c|c|c|c|}
        \hline
        \multicolumn{2}{|c|}{} & \multicolumn{3}{c|}{\textbf{Attack feas. rating}} \\
        \cline{3-5}
        \multicolumn{2}{|c|}{} & Unlikely & Possible & Likely \\
        \hline
        \multirow{4}{*}{\textbf{Impact rating}} 
        & Very High & 3 & 4 & 5 \\
        \cline{2-5}
        & High & 2 & 3 & 4 \\
        \cline{2-5}
        & Medium & 2 & 2 & 3 \\
        \cline{2-5}
        & Low & 1 & 2 & 2 \\
        \cline{2-5}
        & Very Low & 1 & 1 & 1 \\
        \hline
    \end{tabular}
    \label{tab:risk_model}
\end{table}
 
This matrix allows the identified CAPECs to be classified based on severity and likelihood of exploitation, ensuring a direct correspondence with the proposed risk assessment model. Below, Table \ref{tab:capecs_att} presents the relevant attributes of the selected CAPECs, along with their associated impact rating.  
 
\begin{table}[ht]
    \centering
    \caption{\textbf{Relevant attributes of the selected CAPECs.}}
    \begin{tabular}{|p{3.0cm}|p{2.7cm}||p{3.4cm}|p{2.6cm}|||p{1.3cm}|}
        \hline
         \textbf{Attack pattern (CAPEC)} & \textbf{Severity \newline {\scriptsize (Typical\_Severity)}} & \textbf{Attack likel. \newline {\scriptsize (Likelihood\_of\_Attack)}} & \textbf{Expertise \newline {\scriptsize (Skills\_Required)}} & \textbf{Impact \newline rating} \\
        \hline
        668: KNOB & High & Low & Medium & 3 \\
        \hline
        666: BlueSmacking & Medium & Medium & Low & 2 \\
        \hline
    \end{tabular}
    \label{tab:capecs_att}
\end{table}
 
When comparing the values obtained with the examples provided in Annex H of ISO 21434, a slight divergence can be observed in the first scenario evaluated. In the case of the Bluetooth signal spoofing attack (CAPEC-668), the proposed model assigned a risk value of 3, while the evaluation of the standard's example indicated a value of 4. This difference can be attributed to the way attack feasibility is calculated in the proposed approach, which considers attributes such as the required skill level and the likelihood of exploitation. This difference may reflect a more realistic scenario, considering the specific characteristics of the Bluetooth attack, such as the manipulation of authentication keys.
 
In the second scenario, regarding the denial-of-service attack (CAPEC-666), the results were identical in both methods, with both models classifying the risk as 2. This alignment reinforces the consistency of the proposed approach and its ability to maintain compatibility with already consolidated models, such as that of ISO 21434, while providing a higher level of technical detail due to the use of attack-related databases.
 
These results indicate that, although small variations may occur in specific cases, the proposed model is able to adequately reflect the severity and feasibility of the attacks, maintaining adherence to normative principles and adding value through an analysis more oriented toward the real technical scenario of the evaluated system.

\section{Conclusion} \label{sec:conclusion} 
 
This work presented a methodological approach to integrate the STRIDE threat model with the CAPEC database, focusing on risk analysis in automotive systems, in accordance with the guidelines of the ISO/SAE 21434 standard. Through the construction of a MongoDB database and the structured modeling of the relationships between assets, threats, and attack patterns, it was possible to systematically represent vulnerability exploitation paths, enabling the construction of attack trees and the classification of threats in terms of feasibility and impact.
 
The results demonstrate that the association between STRIDE and CAPEC enriches the threat identification and assessment process, providing a solid basis for decisions related to risk mitigation in connected automotive environments. The practical application of this methodology was exemplified through the analysis of threats involving the Bluetooth protocol, highlighting the usefulness of the proposal for real-world scenarios.
 
As future work, we intend to expand the integration of the model with the CWE (\textit{Common Weakness Enumeration}) and CVE (\textit{Common Vulnerabilities and Exposures}) databases. The inclusion of CWE will enable the mapping of software weaknesses related to the attack patterns already identified, deepening the analysis of the origin of vulnerabilities. In turn, the incorporation of CVE will enable direct association with real, documented vulnerabilities, bringing a more up-to-date perspective oriented toward the practical application of automotive cybersecurity. This evolution will contribute to a more complete and effective framework for proactive threat management. 

\appendix

\section{Object-Oriented Data Model} \label{appendix:model}
 
With the objective of integrating CAPEC attack patterns into the ISO/SAE 21434 TARA analysis process, an object-oriented UML data model was developed. The model was built from the structure made available by MITRE, aiming to organize the data systematically to facilitate queries.
 
\begin{figure}[ht]
\centering
\includegraphics[width=1.0\textwidth, page=1, trim={4.2cm 2.9cm 4.1cm 2.4cm},clip]{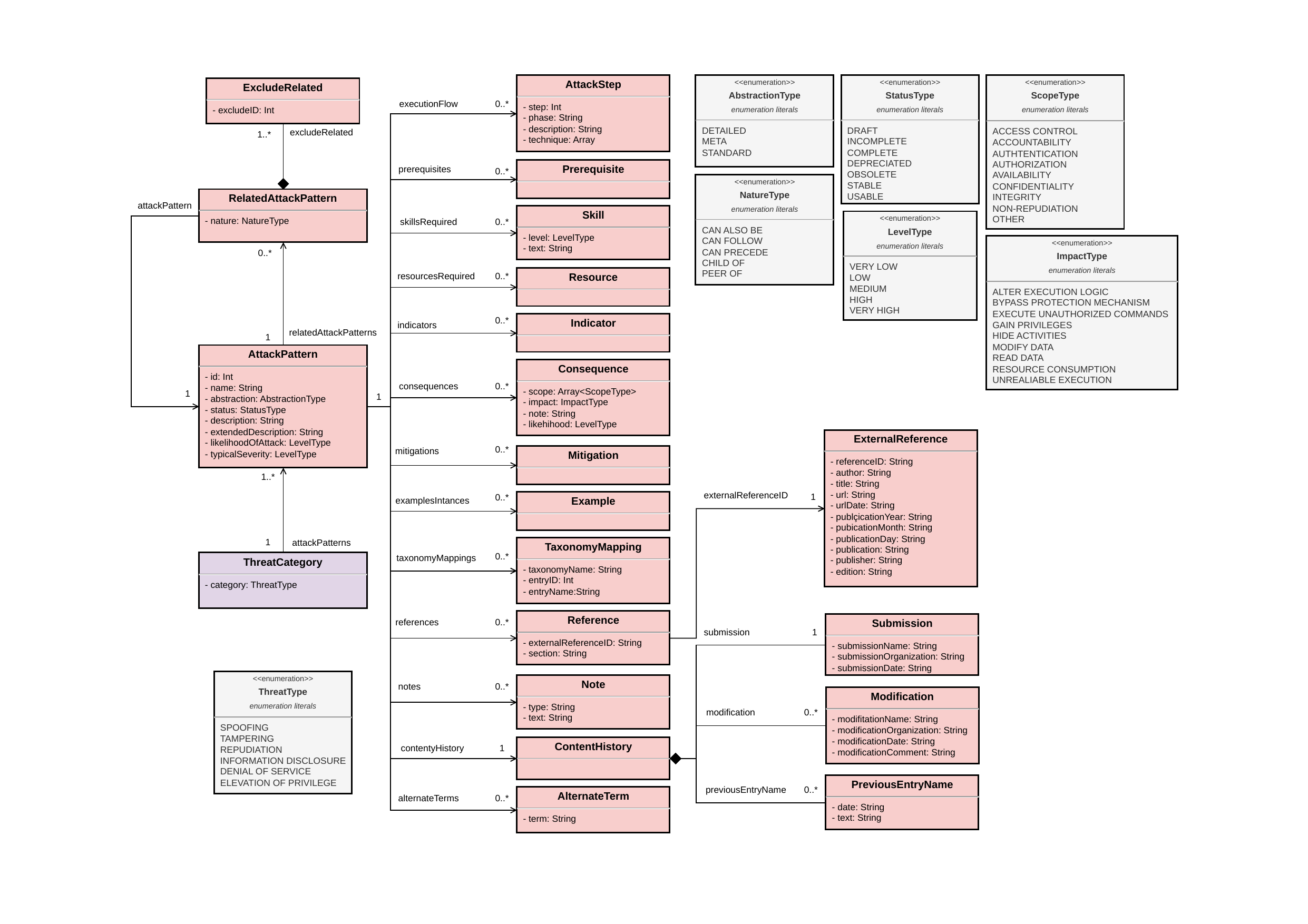}
\caption{Data model defining the relationships between CAPEC and STRIDE.}
\label{fig:CAPEC_model}
\end{figure}
 
Figure~\ref{fig:CAPEC_model} presents the structure of the model, which includes both the entities specific to CAPEC attack patterns and the threat categories of the STRIDE model, allowing a direct association between the two domains. Each CAPEC pattern contains a series of relationships that aim to facilitate its understanding, application, and categorization. Among the entities most relevant to risk assessment are:
 
\begin{itemize}
    \item \textit{executionFlow}: describes, in a sequential and structured way, the \textit{AttackStep}s, defined as the steps necessary for the attack to be carried out. This includes preliminary actions, conditions necessary for the attack's success, and the expected effects on the target system. This description enables the construction of attack trees and an in-depth understanding of the vectors exploited by the adversary; \\[-0.3cm]
    
    \item \textit{typicalSeverity}: represents an estimate of the impact that the attack can cause on a typical system, considering variables such as loss of integrity, availability, and confidentiality. This attribute is expressed in qualitative levels (\textit{Very Low}, \textit{Low}, \textit{Medium}, \textit{High}, and \textit{Very High}), and can be associated with risk metrics used in standards such as ISO 21434; \\[-0.3cm]
    
    \item \textit{skillsRequired}: indicates the level of technical skill an attacker needs to have to successfully execute the attack pattern. Levels are generally classified as \textit{Low}, \textit{Medium}, or \textit{High}, reflecting the complexity of the attack and its accessibility to different types of agents; \\[-0.3cm]
    
    \item \textit{likelihoodOfAttack}: estimates the probability of the attack occurring in a real environment. Like the other attributes, it is also categorized into qualitative levels. This value is important for composing the attack feasibility rating when used in risk matrices.
\end{itemize}
 
The combination of these entities provides essential support for risk prioritization and the definition of mitigating measures in the secure development lifecycle, particularly in embedded automotive systems, where reliability and safety are critical factors. The use of DBSchema facilitated the visualization and validation of the relationships between entities, and also enables future expansions of the model to encompass CWE and CVE data, as proposed in the future work of this study.

%
%
\begin{credits}
\subsubsection{\ackname}
This study was financed in part by the Coordenação de Aperfeiçoamento de Pessoal de Nível Superior -- Brasil (CAPES) -- Finance Code 001. The authors would also like to thank the Federal University of Juiz de Fora (UFJF) for the institutional and scientific support provided throughout this research.

\subsubsection{\discintname}
The authors have no competing interests to declare that are
relevant to the content of this article.
\end{credits}
%
%
\bibliographystyle{splncs04}
\bibliography{mybibliography}
\end{document}